\documentclass[aps,prl,twocolumn,notitlepage,showpacs]{revtex4-1}

\usepackage{amsmath}
\usepackage{amssymb}
\usepackage{graphicx}
\usepackage[colorlinks=true]{hyperref}
\usepackage{bm}
\usepackage{subfigure}
\usepackage{xcolor}

\begin{document}

\title{Electronic Manipulation of Magnon Topology by Chirality Injection from Boundaries}

\author{Seunghun Lee}
\affiliation{Department of Physics, Korea Advanced Institute of Science and Technology, Daejeon 34141, Republic of Korea}
\author{Gyungchoon Go}
\affiliation{Department of Physics, Korea Advanced Institute of Science and Technology, Daejeon 34141, Republic of Korea}
\author{Se Kwon Kim}
\affiliation{Department of Physics, Korea Advanced Institute of Science and Technology, Daejeon 34141, Republic of Korea}

\begin{abstract}
Magnon bands are known to exhibit nontrivial topology in ordered magnets under suitable conditions, engendering topological phases referred to as magnonic topological insulators. Conventional methods to drive a magnonic topological phase transition are bulk magnetic or thermal operations such as changing the direction of an external magnetic field or varying the temperature of the system, which are undesired in device applications of magnon topology. In this work, we lift the limitation of the magnon topology control on the bulk non-electronic manipulation by proposing a scheme to manipulate magnonic topological phases by electronic boundary operations of spin chirality injection. More specifically, we consider a ferromagnetic honeycomb lattice and show that a finite spin chirality injected from the boundary of the system via the spin Hall effects introduces a tunable sublattice-symmetry-breaking mass term to the bosonic counterpart of the Haldane model for the Chern insulators and thereby allows us to electronically manipulate the bulk topology of magnons from the boundary. The ``shoulder'' in the thermal Hall conductivity profile is proposed as an experimental probe of the chirality-induced topological phase transition. The scheme for the boundary manipulation of the magnon topology is shown to work for a honeycomb antiferromagnet as well. We envisage that the interfacial chirality injection may offer a nonintrusive electronic means to tune the static and the dynamical bulk properties of general magnetic systems.
\end{abstract}

\maketitle

\emph{Introduction.}|
Since the experimental discovery of the integer quantum Hall effect~\cite{klitzing_new_1980} and the ensuing extensive studies on topological materials~\cite{hasan_colloquium_2010, qi_topological_2011, haldane_model_1988}, the idea of topology has been established as a powerful tool in condensed matter physics to understand exotic phases of matter and a promising resource for technical applications. Well-known electronic topological phases include a 
quantum spin Hall system realized in the HgTe quantum well~\cite{bernevig_quantum_2006} and a three-dimensional topological insulator Bi$_2$Te$_3$~\cite{chen_experimental_2009}. These electronic topological materials are of fundamental interest as well as practical significance, for they can serve as excellent spin-current sources, enabling the efficient manipulation of spin-based devices~\cite{MellnikNature2014, FanNM2014, WangPRL2015-4, WangNC2017}.

While the above works are for electrons, there has been emerging attention to the topological phases of bosonic systems. In particular, the topological properties of bosonic collective excitations in ordered magnets, called magnons, have drawn physicists' interest in that they can be easily manipulated through magnetic fields and spin torques to investigate their topological properties. In certain magnetic systems referred to as magnonic topological insulators, magnon bands are known to exhibit nontrivial topologies~\cite{ZhangPRB2013, shindou_topological_2013, MookPRB2014, MookPRB2014-2, WangJAP2021}. The magnonic topological insulators have been identified in various setups including honeycomb ferromagnets with magnon-magnon interaction~\cite{lu_topological_2021, MookPRX2021} and magnetoelastic interaction~\cite{go_topological_2019}.

\begin{figure}[]
\includegraphics[width=\columnwidth]{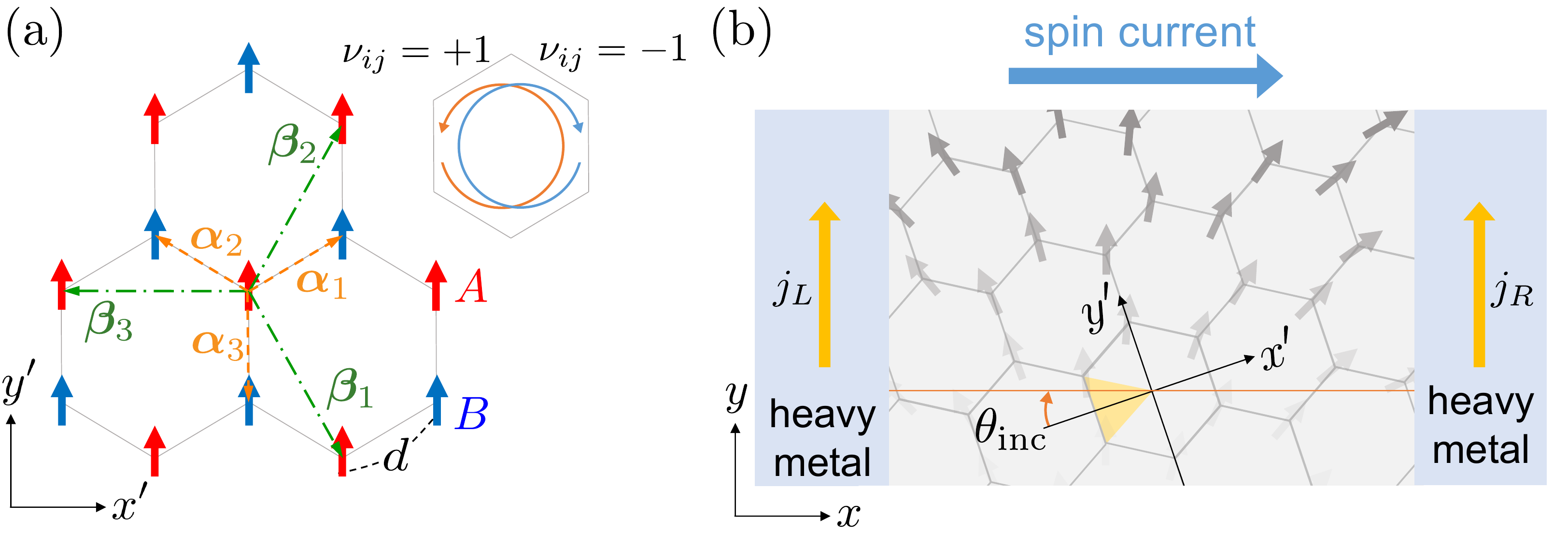}
\caption{(a) The ground-state spin configuration of a ferromagnetic honeycomb lattice and the sign $\nu_{ij}$ of the DM interaction. (b) The chiral spin texture is induced in a honeycomb ferromagnet by injecting a spin current into the sample with the incident angle $\theta_{\text{inc}}$ via the spin Hall effect from the proximate heavy metals. The magnon bands on top of the induced chiral texture can exhibit a topological phase transition as the injected chirality increases. See the main text for the details.}
\label{fig:fig1}
\end{figure}

To harness the magnon topology for practical applications, it is important to be able to manipulate the topological properties of the magnon bands. So far, the conventional control of the magnon-band topology required the bulk manipulation of the systems via, e.g., changing an external magnetic field or the temperature~\cite{KimPRL2019}. However, these bulk magnetic or thermal operations are not only challenging to efficiently implement in nanoscale devices but also unsuitable for ultrafast execution. In this Letter, we lift this limitation of the magnon-band topology control on the bulk non-electronic manipulation by showing that the magnon-band topology can be controlled electrically from the boundary by injecting a spin chirality into the bulk. More specifically, we propose a scheme to manipulate magnonic topological phases of ferromagnets and antiferromagnets on a honeycomb lattice via chirality injection, which can be realized, e.g., via interfacial spin Hall effects resulting from the proximate heavy metals~\cite{TakeiPRL2014, takei_superfluid_2014, takei_nonlocal_2015}. We demonstrate that the chirality injection engenders a sublattice-symmetry-breaking mass on top of the previously studied bosonic counterpart of the Haldane model~\cite{kim_realization_2016, OwerreJPCM2016, OwerreJAP2016}. 
This chirality-induced mass term competes with the known Haldane mass term, which allows us to nonlocally tune the bulk topology from nontrivial to trivial and vice versa from the boundary. For experimental probes of the chirality-induced topological phase transition, we propose the magnon thermal Hall effect arising from the band topology~\cite{matsumoto_theoretical_2011}. The “shoulders” in the thermal Hall conductivity profile can herald the presence of the proposed phase boundary. Our work exemplifies the utility of the interfacial chirality injection via the spin Hall effects as an electronic nonintrusive means to control magnets and magnon bands therein.

\emph{Ferromagnetic Model.}|
We consider a 2D ferromagnet on a honeycomb lattice whose Hamiltonian is given by
\begin{align}\label{Ham}
    H =& -J\sum_{\langle i,j\rangle} \mathbf{S}_i \cdot \mathbf{S}_j + \frac K2 \sum_i S_{i,z}^2 \nonumber \\
    & +D\sum_{\langle\langle i,j \rangle\rangle} \nu_{ij} \hat{\mathbf{z}} \cdot (\mathbf{S}_i \times \mathbf{S}_j) - B\sum_i S_{i,z} \, ,
\end{align}
where the first and the second terms represent the isotropic Heisenberg exchange interaction ($J>0$) between the nearest spins and the easy-plane anisotropy ($K>0$), respectively. The 2D honeycomb ferromagnet with an in-plane magnetic order is realized in monolayer $\text{CrCl}_3$~\cite{lu_meron-like_2020, xue_two-dimensional_2019, DupontPRL2021}. The third term is the Dzyaloshinskii-Moriya (DM) interaction between the next-nearest neighboring spins~\cite{dzyaloshinsky_thermodynamic_1958, KvashninPRB2020}. The DM vector $D_{ij} = D\hat{\mathbf{z}}$ does not have an in-plane component due to the mirror symmetry, and the sign $\nu_{ij} = \pm 1$ of the DM interaction depends on the orientation of the two next-nearest spins as shown in Fig.~\ref{fig:fig1}(a). The last term in Eq.~\eqref{Ham} represents the Zeeman coupling to the external out-of-plane magnetic field $B$. When $|B| < KS$, the ferromagnetic ground state has a uniform spin configuration characterized by a polar angle $\theta_0 = \cos^{-1}(B/KS)$ with the spontaneously broken U(1) spin-rotational symmetry about the $z$ axis, which is crucial for a spin current to flow through the magnet~\cite{SoninJETP1978, KonigPRL2001, sonin_spin_2010, TakeiPRL2014}. We denote the distance between the nearest neighbors on the honeycomb lattice by $d$ and that between the next-nearest neighbors by $a=\sqrt{3}d$. 

Now, let us consider the setup shown in Fig.~\ref{fig:fig1}(b) designed for the injection of spin chirality into the magnet. The honeycomb ferromagnet is sandwiched between two heavy metals with the relative angle of $\theta_{\text{inc}}$ ($|\theta_{\text{inc}}| \leq \pi/6$). The charge currents through the two heavy metals inject a spin current into the ferromagnet via the spin Hall effect~\cite{takei_superfluid_2014, SinovaRMP2015}, which is also referred to as spin-orbit torque~\cite{ManchonRMP2019}. When the charge currents on both sides are the same ($j_L=j_R=j$), which is the situation we focus on in this Letter, it has been shown in Refs.~\cite{TakeiPRL2014, takei_superfluid_2014, takei_nonlocal_2015} that a static spin texture with spatially varying azimuthal angle $\phi(\mathbf{r}) = \phi_0 + \phi' x$ can be established in the bulk of the magnet with the azimuthal-angle gradient $\phi' \approx j \vartheta / A$~\footnote{The Supplemental Material includes the derivation of the continuum Hamiltonian of a honeycomb ferromagnet, the discussion of the system’s capability to host spin supercurrent, the derivation of the equilibrium polar angles of the spins, the perturbative calculation result of the shifted Dirac points and the Dirac Hamiltonians there, and the magnon Hamiltonian of a honeycomb antiferromagnet.}, where $\vartheta$ is the damping-like spin-orbit-torque parameter for the injected spin current per unit charge current density~\cite{TserkovnyakPRB2014}. In this case, the spin current flows with the incident angle of $\theta_{\text{inc}}$ with respect to the primed coordinate system $(x', y')$ of the honeycomb lattice shown in Fig.~\ref{fig:fig1}(b).  Here, note that the spin chirality $\phi'$ of the magnet can be non-locally tuned from the boundary of the sample by controlling the charge currents $j$ flowing in the adjacent heavy metals. Hereafter, we shall use the spin chirality $\phi'$ as a tunable control parameter of the system. When there is a finite spin chirality, the easy-plane anisotropy $K$ is effectively renormalized to $K_{\text{eff}} = K - J[3-g(\phi'd, \theta_{\text{inc}})]$, where $g(\phi'd, \theta_{\text{inc}}) = \sum_{j=1}^3 \cos(\nabla \phi \cdot \bm{\alpha}_j)$ and $\nabla \phi = \phi' (\cos\theta_{\text{inc}}, -\sin\theta_{\text{inc}})$ so that the ferromagnetic ground state has a modified polar angle $\theta_0 = \cos^{-1}(B/K_{\text{eff}} S)$ when $|B| < K_{\text{eff}} S$~\cite{Note1}.

\emph{Magnon Hamiltonian.}|
Under the chirality injection, the static steady state in the bulk of the ferromagnet is given by $\mathbf{S}_i = S(\sin\theta_0 \cos\phi_i, \sin\theta_0 \sin\phi_i, \cos\theta_0)$ where the azimuthal angle varies over space as $\phi_i = \phi_0 + \phi' x_i$. To obtain a magnon band on top of the chiral spin texture, we introduce new spin variables $\tilde{\mathbf{S}}_i = R_i \mathbf{S}_i$ for each spin where the three-dimensional orthonormal matrix $R_i$ represents a local spin transformation that rotates the original spin $\mathbf{S}_i$ to the $z$ axis~\footnote{Explicitly, $R_i = ((\cos\theta_0 \cos\phi_i,  \cos\theta_0 \sin\phi_i,  -\sin\theta_0), \\ (-\sin\phi_i,  \cos\phi_i, 0), (\sin\theta_0 \cos\phi_i, \sin\theta_0 \sin\phi_i, \cos\theta_0))$.}. The Hamiltonian for the noninteracting magnons can be obtained by rearranging Eq.~\eqref{Ham} with new spin variables $\tilde{\mathbf{S}}_i$ and performing the Holstein-Primakoff transformation truncated to the order of $\sqrt{S}$~\cite{holstein_field_1940}: $\tilde{S}_{i,x} = \sqrt{S/2}(c_i^\dagger + c_i), \tilde{S}_{i,y} = i\sqrt{S/2}(c_i^\dagger - c_i)$, and $\tilde{S}_{i,z} = S - c_i^\dagger c_i$ $(c=a,b)$ where $a_i$ and $b_i$ are the independent bosonic operators residing in the sublattice A and B, respectively. After a Fourier transformation, we can rewrite the magnon Hamiltonian in the momentum space in the Bogoliubov-de Gennes (BdG) form~\cite{Note1}. After neglecting the off-diagonal blocks of the BdG magnon Hamiltonian that modify a magnon band structure near the points $\mathbf{K} = (4\pi/3a,0)$ and $\mathbf{K}' = (2\pi/3a,2\pi/\sqrt{3}a)$ only slightly~\footnote{The ratio of the off-diagonal blocks to the diagonal blocks of the BdG Hamiltonian is approximately given by $K/J \ll 1$ and thus, by treating the off-diagonal blocks as perturbations to the diagonal blocks, the correction to the magnon bands is on the order of $K^2 / J \ll K$, which can be neglected compared to the unperturbed magnon gap at the Dirac points.}, we obtain the following magnon Hamiltonian: $H = \sum_{\mathbf{k}} \Psi_{\mathbf{k}}^\dagger [h_0 I + \mathbf{h}(\mathbf{k})\cdot \bm{\sigma}] \Psi_{\mathbf{k}}$ where $\Psi_{\mathbf{k}} = (a_{\mathbf{k}}, b_{\mathbf{k}})^T$ and $\bm{\sigma} = (\sigma_1,\sigma_2,\sigma_3)$ are the Pauli matrices. The components of the Hamiltonian $H$ are given by 
\begin{align}
    h_0 &= \epsilon_0 + JS\sin^2 \theta_0 \sum_{j=1}^3 \cos\varphi_{\alpha_j}, \label{h0} \\ 
    \mathbf{h}(\mathbf{k}) &= \sum_{j=1}^3 \begin{pmatrix}
    -JS \,\text{Re}[\rho_j(\mathbf{k})] \\-JS \,\text{Im}[\rho_j(\mathbf{k})] \\ DS \,f_j(\mathbf{k})
    \end{pmatrix}, \label{h123}
\end{align}
where $\epsilon_0 = 3JS\cos^2 \theta_0 + B\cos\theta_0 - KS(1+3\cos 2\theta_0)/4$, $\rho_j(\mathbf{k}) = e^{-i\mathbf{k} \cdot \bm{\alpha}_j} \left[\cos(\varphi_{\alpha_j}/2) - i\cos\theta_0 \sin(\varphi_{\alpha_j}/2) \right]^2$, $f_j(\mathbf{k}) = \sin\varphi_{\beta_j}\left[ \cos(\mathbf{k}\cdot \mathbf{\beta}_j)(1+\cos^2 \theta_0) - 2\sin^2 \theta_0 \right] + 2\cos\varphi_{\beta_j} \sin(\mathbf{k}\cdot \mathbf{\beta}_j) \cos\theta_0$, $\varphi_{\alpha_j}\equiv \nabla \phi \cdot \bm{\alpha}_j$ and $\varphi_{\beta}\equiv \nabla \phi \cdot \bm{\beta}_j$. The corresponding energies of the upper and the lower bands are given by 
\begin{align}\label{Epm}
    E^{\pm}(\mathbf{k}) = h_0 \pm |\mathbf{h}(\mathbf{k})|.
\end{align}
In the absence of the DM interaction ($D = 0$), the gap between the upper and the lower band closes at the points $\mathbf{k} = \bar{\mathbf{K}}$ and $\bar{\mathbf{K}}'$ where $\rho_j(\mathbf{k}) = 0$ holds, which are so-called the Dirac points. When no spin chirality is present ($\phi' = 0$), the two Dirac points are just $\mathbf{K}$ and $\mathbf{K}'$ and the gap there is given by $\Delta_{\phi'=0} = 6\sqrt{3} DS\cos\theta_0$~\cite{kim_realization_2016, OwerreJPCM2016}. The gap $\Delta_{\phi'=0}$ closes when the DM interaction vanishes.

In the presence of spin chirality $\phi' \neq 0$, the two Dirac points $\bar{\mathbf{K}}$ and $\bar{\mathbf{K}}'$ shift from their original positions and the gap sizes at the two points change. The positions of the shifted Dirac points can be obtained by solving $\rho_j(\mathbf{k}) = 0$ perturbatively with respect to $\phi'$~\cite{Note1}. We can obtain the Dirac Hamiltonian for magnons by expanding the Hamiltonian Eq.~\eqref{h123} near $\bar{\mathbf{K}}$ and $\bar{\mathbf{K}}'$. The size of the modified gaps at $\bar{\mathbf{K}}$ and $\bar{\mathbf{K}}'$ are given by
\begin{align}\label{Gap}
    \Delta_{\bar{\mathbf{K}}/\bar{\mathbf{K}}'} = \frac{3\sqrt{3} |D|}{16K^2} \big|&\pm 8B(4K+3J(\phi'd)^2) \nonumber \\
    & - 13K^2 S (\phi' d)^3 \cos 3\theta_{\text{inc}} \big| \, ,
\end{align}
to linear order in $B$. Note that the gap $\Delta_{\bar{\mathbf{K}}/\bar{\mathbf{K}}'}$ is proportional to the DMI coefficient $D$ as previously shown~\cite{kim_realization_2016} and also proportional to $\cos (3 \theta_\text{inc})$ as dictated by the sixfold rotational symmetry of the honeycomb lattice. More importantly, for $D \neq 0$, there is a critical magnetic field $B_c$ where the gap at either $\bar{\mathbf{K}}$ or $\bar{\mathbf{K}}'$ closes:
\begin{align}\label{Bc}
    B_{\bar{\mathbf{K}}/\bar{\mathbf{K}}',c} = \pm \frac{13KS}{32} (\phi'd)^3 \cos 3\theta_{\text{inc}} \, ,
\end{align}
to cubic order in the spin chirality $\phi 'd$. The critical magnetic field is determined by the spin chirality $\phi'$ and thus can be controlled from the boundaries. This chirality-induced magnon-gap closing at one of the Dirac points is one of our main results, whose topological significance is described below.

\begin{figure}[]
\includegraphics[width=\columnwidth]{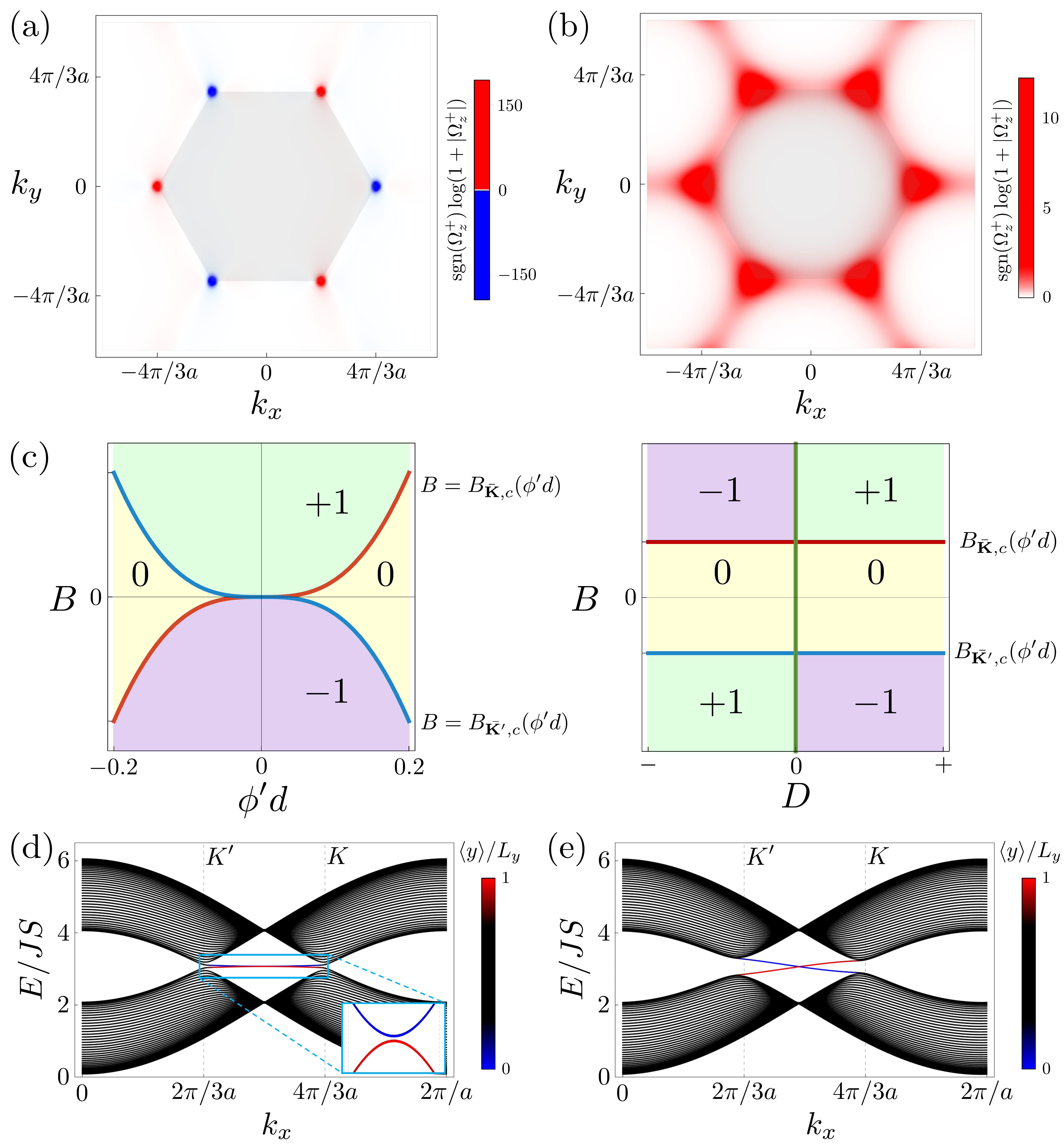}
\caption{(a) The Berry curvature $\Omega_z^+$ of the upper magnon band in a honeycomb ferromagnet under chirality injection $\phi' d = 0.2$ for $B=0$ and (b) for $B=0.05JS$. (c) The topological phase diagram for a ferromagnet, which shows the Chern number of the upper magnon band. (Left panel: $D>0$ is fixed, right panel: $\phi'd>0$ is fixed.) The red and blue curves show the phase boundaries between the trivial and nontrivial phases [Eq.~\eqref{Bc}], and the green line represents the $D=0$ line. (d) The magnon bands in a ferromagnet with ribbon geometry with 40 unit cells for $B=0$ and (e) for $B=0.05JS$. The colors of the bands represent the average vertical positions $\langle y \rangle$ of each mode along the axis where the sample is finite. The inset in (d) zooms in the trivial edge modes. For (a)-(b) and (d)-(e), the parameters $K=D=0.2J, \phi'd=0.2$, and $\theta_{\text{inc}} = 0$ are used.}
\label{fig:fig2}
\end{figure}

\emph{Topological Phase Transition.}|
The band topology of the bosonic quadratic Hamiltonian [Eq.~\eqref{h123}] is identical to that of the corresponding fermionic counterpart with the same Hamiltonian~\cite{chaudhary_simple_2021}. Note from Eq.~\eqref{Gap} that the mass term of the Dirac Hamiltonian [Eq.~\eqref{h123}] contains two components. The first component is the Haldane mass~\cite{haldane_model_1988} proportional to $DB$ and its sign is different between the two Dirac points $\bar{\mathbf{K}}$ and $\bar{\mathbf{K}}'$. This manifestation of the Haldane model at the honeycomb ferromagnet results in topologically nontrivial phases with the Chern numbers $\text{sgn}(DB)$ for the upper magnon band~\cite{kim_realization_2016}. The second component proportional to $(\phi'd)^3$ is a sublattice-symmetry-breaking mass that acts as a staggered on-site potential. The Haldane mass and the sublattice-symmetry-breaking mass compete to determine whether the phase is topologically trivial or not. Since the sublattice-symmetry-breaking mass depends on the spin chirality $\phi'$ injected from the boundaries, we can non-locally manipulate the topological phase of the honeycomb ferromagnet from the boundaries by controlling charge currents in heavy metals.

The topological phase of our ferromagnetic system can be specified by the Chern numbers defined by $C^{\pm} = (1/2\pi)\int_{\text{BZ}} d^2k \Omega_z^{\pm}$ where $\Omega_z^{\pm}(\mathbf{k}) = \mp \frac 12 \hat{\mathbf{n}} \cdot (\partial_{k_x} \hat{\mathbf{n}} \times \partial_{k_y} \hat{\mathbf{n})}$ is the Berry curvatures of each magnon band and $\hat{\mathbf{n}}(\mathbf{k}) = \mathbf{h}(\mathbf{k}) / |\mathbf{h}(\mathbf{k})|$. Figures~\ref{fig:fig2}(a) and (b) show the Berry curvature of the upper magnon band for the incident angle $\theta_\text{inc} = 0$ for two different values of $B$. When the external field is below the critical value (determined by the spin chirality), $|B| < |B_c|$ [see Fig.~\ref{fig:fig2}(a)], the Chern number $C^+$ for the upper band is zero. On the other hand, for the external field exceeding the critical value, the Chern number becomes $C^+ = +1$ for $B > |B_c|$ [see Fig.~\ref{fig:fig2}(b)] and $-1$ for $B < -|B_c|$. Figure~\ref{fig:fig2}(c) summarizes the topological phases of our system. The phase diagram is truncated at $\phi' d = \pm 0.2$, which is below the Landau criterion for the spin chirality (for the used parameters) over which the static stead state becomes unstable~\cite{sonin_spin_2010}. 

A nontrivial band topology in bulk is followed by the existence of chiral edge modes according to the bulk-boundary correspondence~\cite{hasan_colloquium_2010}. The number of the chiral edge modes is dictated by the Chern number, which is a bulk property. Figures~\ref{fig:fig2}(d) and (e) plots the magnon bands at a ribbon geometry for the magnetic fields corresponding to Fig.~\ref{fig:fig2}(a) and (b). We can see that the localized edge modes in Fig.~\ref{fig:fig2}(d) can be deformed smoothly into bulk bands, while that in Fig.~\ref{fig:fig2}(c) bridge the gap and hence are chiral. The change of the magnon topology driven by the spin chirality $\phi'$ injected from the boundary constitutes one of our main results.

Here, we remark that the effect of the spin chirality on the topological properties of magnon bands depends on the incident angle $\theta_\text{inc}$ through the sublattice-symmetry-breaking mass term $\propto (\phi'd)^3 \cos \theta_\text{inc}$ in Eq.~\eqref{Gap}. In particular, when $\theta_{\text{inc}} = \pm\pi/6$, the sublattice-symmetry-breaking mass term is absent in the magnon Dirac Hamiltonian and thus the magnon band's Chern numbers remain finite regardless of the injected spin chirality. However, as long as $\theta_{\text{inc}} \neq \pm\pi/6$, the system can be driven from the nontrivial phase into the trivial phase and vice versa by non-locally controlling the spin chirality of the sample, providing an example of the bulk topology manipulation from the boundaries.

\begin{figure}[]
\includegraphics[width=0.9\columnwidth]{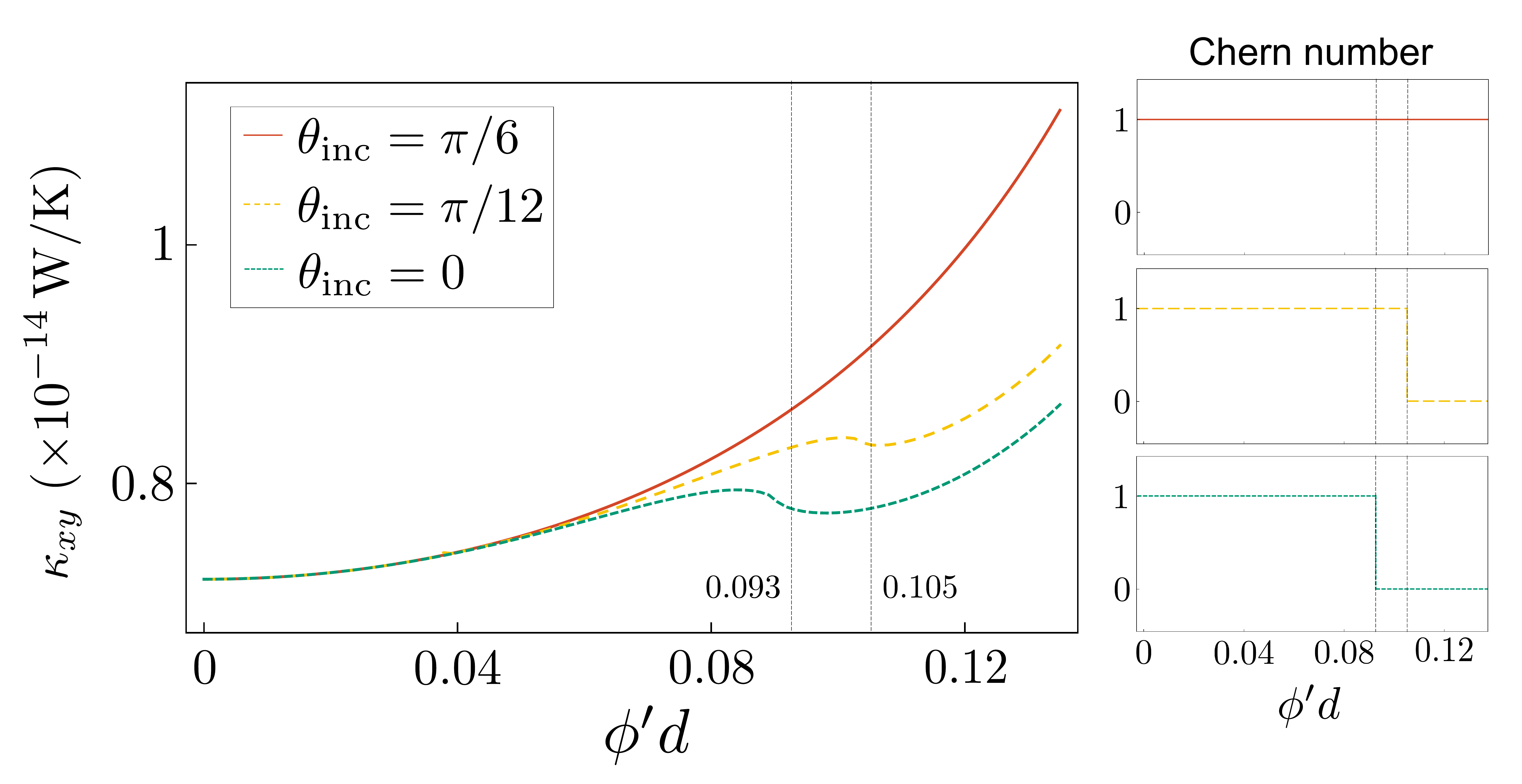}
\caption{Thermal Hall conductivity $\kappa_{xy}$ as a function of the injected spin chirality $\phi'$, where $d$ is the lattice constant. The ``shoulders" appear at the topological phase transition [Eq.~\eqref{Bc}] for the incident angle $\theta_{\text{inc}} = 0$ and $\pi/12$. For the incident angle $\theta_\text{inc} = \pi/6$, there is no topological phase transition and thus no ``shoulders'' pattern in the thermal Hall conductivity. The plots on the right panel show the Chern numbers of the upper band as a function of the spin chirality. The used parameters are given in the main text.}
\label{fig:fig3}
\end{figure}

\emph{Thermal Hall Effect.}|
Our theoretical prediction of the chirality-induced topological phase transition can be experimentally examined by measuring the magnon thermal Hall conductivity, which can be expressed in terms of the Berry curvatures $\Omega_z^\pm$~\cite{matsumoto_theoretical_2011, MurakamiJPSJ2017}:
\begin{align}
    \kappa_{xy} = -\frac{k_B^2 T}{\hbar} \sum_{n=\pm} \int \frac{d^2 \mathbf{k}}{(2\pi)^2}\, c_2(\rho_{n,\mathbf{k}})\, \Omega_z^n (\mathbf{k}) \, ,
\end{align}
where $\rho_{n,\mathbf{k}} = (e^{E_{n,\mathbf{k}}/k_B T} - 1)^{-1}$ is the Bose-Einstein distribution function, $c_2(\rho) = (1+\rho) [\ln((1+\rho)/\rho)]^2 - (\ln\rho)^2 - 2\text{Li}_2 (-\rho)$ and $\text{Li}_2 (z)$ is the polylogarithm function. To estimate the thermal Hall conductivity numerically, we adopt the following material parameters for CrCl$_3$. We assume $S=3/2$, which is a Cr magnetic moment of bulk chromium trihalides \cite{dillon_magnetization_1965}. The exchange and the easy-plane anisotropy coefficients are given by $J = 0.79$ meV and $K = 0.03$ meV~\cite{lu_meron-like_2020}. For the DM interaction, we use $D = 0.2$ meV~\cite{cai_topological_2021}. We choose $B = 2.2$ Oe  for the external magnetic field. With the given parameters, Fig.~\ref{fig:fig3} plots the thermal Hall conductivity $\kappa_{xy}$ at $T = 10$ K as a function of the spin chirality $\phi'$. For the small spin chirality $\phi' d \lesssim 0.06$, the thermal Hall conductivity $\kappa_{xy}$ increases as the spin chirality $\phi'$ increases because of two reasons. First, the constant energy term $h_0$ [Eq.~\eqref{h0}] decreases as the spin chirality increases. This lowers the energy level at Dirac points and $c_2(\rho)$ increases accordingly. Second, the small spin-chirality-induced gap (the second term in Eq.~\eqref{Gap}) broadens the gap width so that a broader region near the Dirac points contributes to $\Omega^n_z(\mathbf{k})$. When $\theta_{\text{inc}} \neq \pm\pi/6$, the profile of the thermal Hall conductivity develops a ``shoulder" as we pass through the topological phase transition [Eq.~\eqref{Bc}]. These shoulders arise due to the change of the Chern number from $\pm1$ to 0 at the phase boundary. Away from the phase boundary, the thermal Hall conductivity begins to increase again. When $\theta_{\text{inc}} = \pm \pi/6$ for which no topological phase transition occurs, the Chern number does not change and hence there is no ``shoulder'' pattern in the thermal Hall conductivity. Detecting the ``shoulder'' patterns in the profile of $\kappa_{xy}$ can provide the feasibility of the suggested manipulation of magnonic topological phases.

\emph{Discussions.}|
In this Letter, we have investigated the manipulation of the magnon topology of honeycomb ferromagnets via the interfacial chirality injection. We have shown that the bulk magnon topology of the magnets can be electronically manipulated by inducing the spin chirality from the boundaries through the charge currents in the proximate heavy metals. As an experimental probe of the predicted chirality-induced topological phase transition of magnons, we have proposed the shoulder-like feature of the magnon thermal Hall conductivity. In this work, the effect of the charge currents in the proximate heavy metals is taken into account only through the induction of a static chiral spin texture in the magnet. While a finite non-equilibrium coupling is expected to exist between the charge currents and magnon edge modes via the interfacial spin Hall effect, it would not affect the topological property of the bulk magnon bands that is determined by the background spin configuration.

We remark that the predicted topological phase transition occurs also in a honeycomb antiferromagnet whose Hamiltonian is given by Eq.~\eqref{Ham} with $-J$ replaced by $J>0$. Analogous to the ferromagnetic case, the spin chirality can be injected into the antiferromagnet by sandwiching it with heavy metals and flowing parallel currents through the heavy metals~\cite{takei_superfluid_2014, YuanSA2018}. It can be shown that the injected spin chirality can drive a topological phase transition of magnons in antiferromagnetic honeycombs subjected to an external field. The detailed discussion of the antiferromagnetic case is in the Supplemental Material~\cite{Note1}. Our results indicate that the interfacial chirality injection might serve as versatile nonintrusive means to control the static and dynamical bulk properties of generic magnetic systems.

\begin{acknowledgments}
We acknowledge the useful discussion with Yaroslav Tserkovnyak. This work was supported by Brain Pool Plus Program through the National Research Foundation of Korea funded by the Ministry of Science and ICT (NRF-2020H1D3A2A03099291), by the National Research Foundation of Korea(NRF) grant funded by the Korea government(MSIT) (NRF-2021R1C1C1006273), and by the National Research Foundation of Korea funded by the Korea Government via the SRC Center for Quantum Coherence in Condensed Matter (NRF-2016R1A5A1008184). G.G. was supported by the National Research Foundation of Korea (NRF-2022R1C1C2006578).
\end{acknowledgments}

\bibliographystyle{apsrev4-1}
\bibliography{ref, /Users/kimsek/Dropbox/School/Research/master}

\end{document}